\documentclass[sigconf,nonacm]{acmart}
\usepackage{hyperref}
\usepackage[utf8]{inputenc}
\usepackage[T1]{fontenc}
\usepackage{microtype}
\usepackage{xcolor}
\usepackage{xspace}

\let\oldcite\cite
\renewcommand{\cite}[1]{{\color{green!50!black}\oldcite{#1}}}

\makeatletter
\let\origref\ref
\renewcommand{\ref}[1]{%
  \def\@tmpa{#1}%
  \@check@fig@tab{#1}%
}
\newcommand{\@check@fig@tab}[1]{%
  \expandafter\@checkprefix\expandafter{#1}{fig:}{\textcolor{red}{\origref{#1}}}{%
    \expandafter\@checkprefix\expandafter{#1}{tab:}{\textcolor{red}{\origref{#1}}}{%
      \origref{#1}%
    }%
  }%
}
\newcommand{\@checkprefix}[4]{%
  \def\@tmppre{#2}%
  \expandafter\@@checkprefix#1\@nil{#2}{#3}{#4}%
}
\def\@@checkprefix#1:#2\@nil#3#4#5{%
  \def\@tmpkey{#1:}%
  \ifx\@tmpkey\@tmppre #4\else #5\fi
}
\makeatother

\usepackage{graphicx}
\usepackage{textcomp}
\usepackage{threeparttable}
\usepackage{booktabs}
\usepackage{enumitem}
\usepackage{float}
\usepackage{caption}
\usepackage{subcaption}
\usepackage{multirow}
\usepackage{tabularx}
\usepackage{pifont} 

\usepackage[varqu]{zi4}
\usepackage{listings}
\usepackage{tcolorbox}
\tcbuselibrary{listings,skins,breakable}

\definecolor{codebg}{HTML}{F7F7F8}
\definecolor{codeframe}{HTML}{D0D0D0}
\definecolor{diffgreen}{rgb}{0.0, 0.5, 0.0}
\definecolor{diffred}{rgb}{0.6, 0.0, 0.0}
\definecolor{meta}{rgb}{0.0, 0.0, 0.8}
\definecolor{diffaddbg}{HTML}{E6FFEC}
\definecolor{diffrembg}{HTML}{FFEBE9}
\definecolor{keywordblue}{HTML}{0550AE}
\definecolor{stringgreen}{HTML}{0A6640}
\definecolor{commentgray}{HTML}{6A737D}
\definecolor{decoratororange}{HTML}{CF8600}
\definecolor{leftbar}{HTML}{3B82F6}

\lstdefinelanguage{diff}{
    morecomment=[f][\color{meta}]{@@},
    morecomment=[f][\color{gray}]{---},
    morecomment=[f][\color{gray}]{+++},
    morecomment=[f][\color{diffgreen}]{+},
    morecomment=[f][\color{diffred}]{-},
    morecomment=[f][\color{gray}]{diff --git},
    keepspaces=true,
}

\lstdefinestyle{pythonstyle}{
    language=Python,
    keywordstyle=\color{keywordblue}\bfseries,
    stringstyle=\color{stringgreen},
    commentstyle=\color{commentgray}\itshape,
    emph={self,True,False,None},
    emphstyle=\color{keywordblue},
    morekeywords={dataclass,abstractmethod,property},
    literate={->}{{{\color{keywordblue}->}}}2
             {...}{{{\color{commentgray}...}}}3,
}

\newtcblisting{diffcode}[1][]{%
    enhanced,
    listing only,
    listing options={language=diff, basicstyle=\ttfamily\footnotesize, columns=fullflexible, keepspaces=true, breaklines=true},
    colback=codebg,
    colframe=codeframe,
    boxrule=0.4pt,
    arc=2pt,
    left=4pt, right=4pt, top=2pt, bottom=2pt,
    borderline west={2pt}{0pt}{diffred!60},
    #1
}

\newtcblisting{pycode}[1][]{%
    enhanced,
    listing only,
    listing options={style=pythonstyle, basicstyle=\ttfamily\footnotesize, columns=fullflexible, keepspaces=true, breaklines=true},
    colback=codebg,
    colframe=codeframe,
    boxrule=0.4pt,
    arc=2pt,
    left=4pt, right=4pt, top=2pt, bottom=2pt,
    borderline west={2pt}{0pt}{leftbar},
    #1
}

\newtcblisting{celldiff}[1][]{%
    enhanced,
    listing only,
    listing options={language=diff, basicstyle=\ttfamily\scriptsize, columns=fullflexible, keepspaces=true, breaklines=true},
    colback=codebg,
    colframe=codebg,
    boxrule=0pt,
    arc=1pt,
    left=2pt, right=2pt, top=1pt, bottom=1pt,
    #1
}

\newcommand{\sys}{\textsc{PatchAdvisor}\xspace}
\newcommand{\syzbot}{\textsc{syzbot}\xspace}
\newcommand{\nbugs}{6{,}946}
\newcommand{\npatch}{5{,}043}
\newcommand{\ndisc}{5{,}000}

\begin{document}

\title{Beyond Crash-to-Patch: Patch Evolution for Linux Kernel Repair}

\author{Luyao Bai}
\email{lbai27@uic.edu}
\affiliation{\institution{University of Illinois Chicago}\country{USA}}

\author{Kenan Alghythee}
\email{kalghy2@uic.edu}
\affiliation{\institution{University of Illinois Chicago}\country{USA}}

\author{Hang Zhang}
\email{hz64@iu.edu}
\affiliation{\institution{Indiana University}\country{USA}}

\author{Xiaoguang Wang}
\email{xgwang9@uic.edu}
\affiliation{\institution{University of Illinois Chicago}\country{USA}}

\begin{abstract}

Linux kernel bug repair is typically approached as a direct mapping from crash reports to code patches. In practice, however, kernel fixes undergo iterative revision on mailing lists before acceptance, with reviewer feedback shaping correctness, concurrency handling, and API compliance. This iterative refinement process encodes valuable repair knowledge that existing automated approaches overlook.

We present a large-scale study of kernel patch evolution, reconstructing \nbugs{} syzbot-linked bug-fix lifecycles that connect crash reports, reproducers, mailing-list discussions, revision histories, and merged fixes. Our analysis confirms that accepted repairs are frequently non-local and governed by reviewer-enforced constraints not present in bug reports. Building on these insights, we develop \sys{}, a repair framework that integrates retrieval-based memory with a fine-tuned diagnostic advisor to guide a coding agent toward reviewer-aligned patches. Evaluation on temporally held-out syzbot cases demonstrates that leveraging patch-evolution history yields measurable gains in both reviewer-aligned refinement signals and end-to-end repair quality compared to unguided and retrieval-only baselines.

\end{abstract}

\settopmatter{printfolios=true}
\maketitle
\pagestyle{plain}

\section{Introduction}
\label{sec:intro}

Continuous fuzzing infrastructure such as \syzbot{}~\cite{syzkaller,syzbot} has dramatically accelerated the discovery of Linux kernel bugs, yet translating crash reports into correct, merge-worthy patches remains a bottleneck. The difficulty lies not only in generating plausible code changes, but in satisfying the implicit constraints that kernel maintainers enforce during review.

\begin{figure}[htbp]
    \centering
    \includegraphics[width=\linewidth]{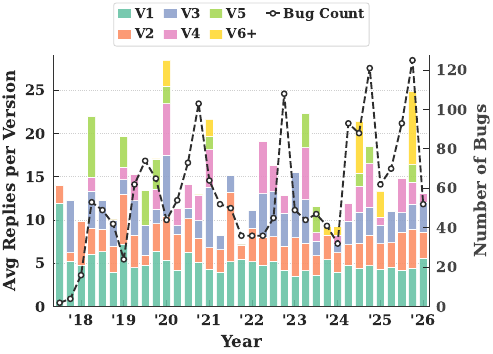}
    \caption{\textit{Review effort across patch versions.} A quarterly stacked area chart showing the average number of discussion replies per patch version for syzbot-reported bugs. The lower layer (V1) represents initial proposals; upper layers (V2--V6+) represent subsequent revisions. The dashed line tracks the quarterly bug count.
    }
    \label{fig:bug_trend}
\end{figure}

In the Linux kernel workflow, candidate fixes are submitted to mailing lists, scrutinized by subsystem experts, and revised---often through multiple patch versions---before acceptance (Figure~\ref{fig:bug_trend}). Each revision incorporates feedback on correctness, locking discipline, API conventions, error handling, and interactions with neighboring subsystems. The final merged patch therefore reflects not just a code change but the outcome of an iterative refinement process whose intermediate reasoning is lost if one examines only the bug report and the eventual fix.

This observation has direct implications for automated program repair (APR)~\cite{GenProg2012,ASRSurvey2018,LLMsSWEReview2024}. Current LLM-based repair methods predominantly treat bug fixing as one-step patch generation from failure evidence~\cite{Zubair-APR-LLM-2025, Jiang-APR-LLM-2024,Yu-APR-LLM-2025,Huang-APR-LLM-2025,Huan-APR-LLM-2023,Kong-APR-LLM-2025,Nong-APR-LLM-2025,Kim-APR-LLM-2025,Seo-APR-LLM-2025,Bouzenia-APR-LLM-2025,Pearce-APR-LLM-2023,Li-APR-LLM-2025,Xia-APR-LLM-2023,Fan-APR-LLM-2023}. For kernel development, this framing is insufficient: a patch that silences a crash may still be rejected for violating subsystem invariants, addressing a symptom rather than the root cause, or introducing regression risk. The challenge is generating patches that not only compile and pass tests, but also survive expert review.

We posit that patch revision history offers a largely untapped supervision signal for this problem. Unlike bug reports and final commits alone, revision chains and review threads reveal \emph{why} an initial candidate fell short, \emph{what} constraints reviewers imposed, and \emph{how} the repair evolved into an acceptable solution. This makes patch evolution both a lens for understanding kernel repair and a concrete source of training signal for AI-assisted patch generation.

To investigate this hypothesis, we reconstruct a dataset of \syzbot-linked Linux kernel bug-fix lifecycles~\cite{syzbot}, connecting crash evidence, reproducer artifacts, mailing-list discussions, revision histories, and merged fixes from multiple public sources. The resulting corpus covers 6,946 fixed bugs, with recovered merged diffs for 5,043 and discussion threads for approximately 5,000. This representation enables studying repair as an iterative process of proposal, critique, and refinement rather than a static input--output mapping.

Our empirical analysis confirms that kernel repair is frequently iterative, non-local, and shaped by reviewer-enforced constraints absent from the original bug report. Accepted changes often occur far from the crash site, and review discussion regularly surfaces additional requirements on concurrency, API usage, and maintainability. These patterns indicate that modeling kernel repair as \emph{one-shot crash-to-patch translation} is fundamentally incomplete.

Building on these findings, we develop \sys{}, a repair framework that incorporates patch-evolution history via two complementary mechanisms: a layered memory that retrieves similar bugs, fix strategies, and review-derived lessons; and a fine-tuned advisor model that predicts likely root-cause directions and reviewer concerns for a given crash. These components jointly guide a general-purpose coding agent toward patches better aligned with historically accepted fixes.
Our evaluation provides initial evidence that evolution-aware guidance improves repair quality. On 100 temporally held-out \syzbot-related patch series, the memory subsystem improves alignment with reviewer-raised revision categories; on recent end-to-end repair cases, the best advisor configuration outperforms both unguided generation and retrieval-only guidance.

In summary, this paper makes the following contributions:

\begin{itemize}[leftmargin=*,nosep]
    \item \textbf{A large-scale dataset of kernel patch evolution.}
    We reconstruct a \syzbot-grounded corpus linking crash reports,
    reproducers, patch discussions, revision histories, and merged
    fixes for nearly 7,000 bugs, enabling fine-grained study of
    Linux kernel repair processes.

    \item \textbf{An empirical characterization of iterative kernel repair.}
    We demonstrate that kernel bug fixing is frequently non-local,
    multi-version, and governed by reviewer-enforced constraints
    invisible in crash reports alone.

    \item \textbf{A proof-of-concept evolution-aware repair system.}
    We present \sys{}, combining retrieval-based memory with a
    fine-tuned diagnostic advisor. Initial results on held-out
    \syzbot{} bugs indicate that patch-evolution guidance can
    improve both refinement signal quality and end-to-end repair
    outcomes.
\end{itemize}

\section{Background and Motivation}
\label{sec:background}

\emph{Kernel Bug Fixing.}
\syzbot{}~\cite{syzbot} is Google's continuous fuzzing platform for the Linux kernel. It systematically exercises upstream kernels~\cite{syzkaller}, detects crashes, minimizes and preserves reproducers, and publishes bug reports through a public dashboard~\cite{syzbot}. For many bugs, \syzbot{} further tracks their lifecycle by linking reports to the commits that resolve them. This makes \syzbot{} a rich source of real-world kernel bug-fixing data---spanning bug reports, reproducers, and corresponding fixes---with over 7,000 \textit{fixed bugs} recorded to date.

Critically, \syzbot{} exposes a bug's \emph{symptoms}---crash reports, stack traces, reproducers---rather than its underlying root cause. Bridging this symptom--cause gap still depends heavily on manual diagnosis and developer expertise~\cite{wu2023klaus,zhang2021androidkernelpatch,mathai2024kgym}.

In contrast to many user-space projects that rely on pull-request-based development, Linux kernel patches are reviewed primarily through email-based discussions on the Linux kernel mailing list (LKML)~\cite{lkml} and subsystem-specific mailing lists~\cite{lore}. A developer typically submits an initial patch (v1), receives comments from maintainers and reviewers, and then posts revised versions (v2, v3, \ldots) until the patch is accepted and merged, or rejected.
This process naturally produces a sequence of patch revisions together with the corresponding review discussion.

Consequently, for Linux kernel bugs, a final fixing commit often reveals only the end result of repair, but not the reasoning that led to it. Earlier patch versions and review feedback capture what was incomplete, incorrect, or unacceptable in an initial fix attempt, as well as the subsystem constraints that shaped later revisions. This revision history therefore provides a richer view of kernel repair than bug reports and final commits alone.

\emph{Motivation.}
The central insight motivating our work is that patch evolution exposes the gap between \textit{a plausible fix} and \textit{a merge-worthy fix}. Reconstructing and analyzing this evolution reveals recurring revision patterns and reviewer constraints, offering a foundation for both understanding and supporting Linux kernel bug repair.

\section{Data Collection and Engineering}
\label{sec:data}

Our first goal is to construct a structured dataset that links kernel bug reports to the full evolution of their fixes. The main challenge is that the relevant artifacts (e.g., bug reports, patch revisions, review discussions, and merged commits) are distributed across multiple data sources~\cite{lkml,syzbot,lore,linux_git,patchwork} and often lack explicit linkage. Recovering a coherent bug-fix history therefore requires substantial data integration and normalization.

To support this effort, we built infrastructure for collecting and parsing a broad kernel developers' email archive~\cite{lkml,lore}. This larger corpus is useful for message lookup, thread recovery, and future expansion beyond a single reporting channel.
However, the dataset used in this paper is intentionally narrower. We focus on bugs reported by \syzbot{}~\cite{syzbot}, because \syzbot{} provides a reliable bug anchor: each report identifies a concrete failure instance and often exposes crash reports, reproducer artifacts, discussion links, and eventual fixing commits. This bug-centered design makes it possible to connect failure evidence with patch discussions and revision history at a higher precision than archive-wide mining alone.

Starting from fixed \syzbot{} reports, we construct the \syzbot{}-grounded dataset through a
multi-stage pipeline that integrates data from the \syzbot{} API,
\texttt{git.kernel.org}, \texttt{lore.kernel.org}, and
\texttt{patchwork.kernel.org}. The resulting dataset links four classes of
artifacts for each bug: crash evidence, reproducer inputs, patch iterations and
discussions, and final merged patch fixes.

\subsection{Broader LKML Archive Collection}

As our first step, we collected a large archive of LKML emails spanning ten years. For each email, we parse metadata such as subject, timestamp, sender, \texttt{Message-ID}, and reply references, together with the message body. This archive gives us an initial understanding of the kernel development process and provides a foundation for email thread lookup and context recovery when direct \syzbot{} links are incomplete.

Mining LKML at scale is complicated by the fact that revised patch versions
are not always posted as strict replies to earlier submissions. Developers
frequently resend updated versions as new top-level emails, sometimes with
modified subject prefixes or subsystem tags. This makes simple
header-following insufficient for reconstructing patch evolution and
motivates the need for normalization and recovery logic.

\subsection{syzbot-Grounded Dataset Construction}

Although we maintain a broader LKML archive, the analysis in this paper is
performed on the subset of repair activity that can be grounded in fixed
\syzbot{} bugs. We begin from the \syzbot{} endpoint for upstream bugs marked
as fixed. Each \syzbot{} entry serves as the root object in our dataset and
provides metadata including the bug title, crash timestamps, discussion
links, and fixing commit hashes.

For each bug, we additionally retrieve up to several crash-related instances,
including the crash report, C reproducer, and syzkaller reproducer program
when available. These artifacts provide concrete failure evidence and help
distinguish one bug from another. In total, we collect \nbugs{} fixed bugs.

\begin{figure*}[t]
    \centering
    \includegraphics[width=\textwidth]{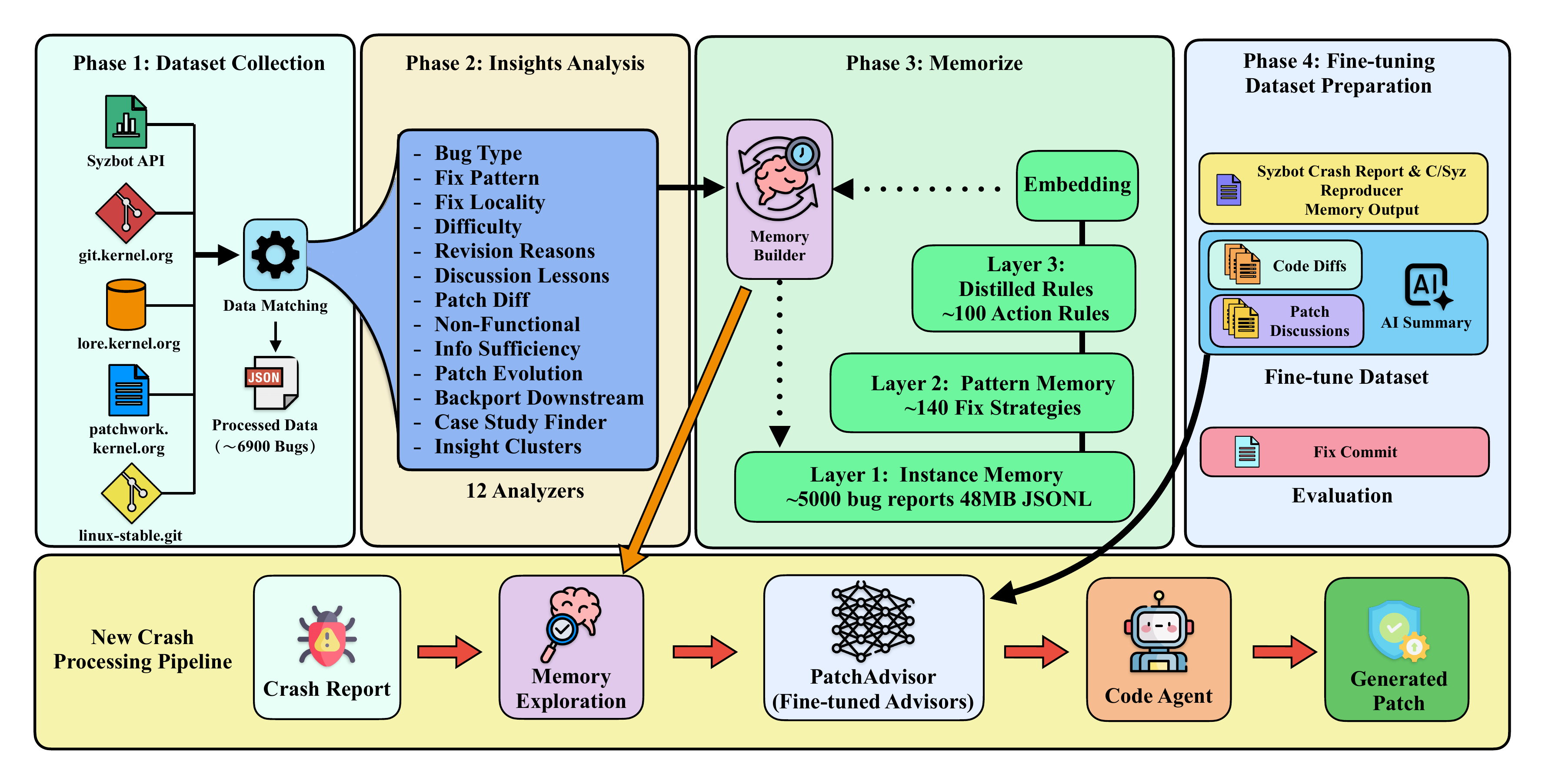}
    \vspace{-8mm}
    \caption{End-to-end pipeline for building \sys from syzbot data: bug reports and patch history are collected and analyzed, compiled into layered memory and training corpora, and then retrieved at inference time to guide LLM-based patch or review generation.}
    \label{fig:data_pipeline}
\end{figure*}

\subsubsection{Repair Artifact Recovery}

For each bug with an associated fixing commit, we retrieve the merged patch diff from \texttt{git.kernel.org}. These diffs represent the final accepted repair state. When a commit is not present in the main \texttt{torvalds/linux} repository, we fall back to subsystem-specific trees such as \texttt{netdev/net}, \texttt{netdev/net-next}, \texttt{bpf/bpf}, and \texttt{bpf/bpf-next}. We successfully recover merged diffs for \npatch{} bugs.
To reconstruct the repair discussion, we follow the discussion links exposed by \syzbot{} and download the corresponding threads from \texttt{lore.kernel.org} as mbox archives~\cite{loredocs}. We parse each thread into individual messages and preserve metadata including sender, subject, timestamp, \texttt{Message-ID}, and \texttt{In-Reply-To}. These metadata are used to recover the structure of patch submission and review discussions.

A primary goal of this stage is to reconstruct patch revision history. We
identify version markers such as \texttt{[PATCH v2]} and
\texttt{[PATCH v3 1/3]} from subject lines and group related messages by
revision. When the links exposed by \syzbot{} are incomplete, we search
\texttt{lore.kernel.org} by patch title and supplement missing information
from \texttt{patchwork.kernel.org}. This recovery process improves coverage
of patch discussions and helps us connect earlier candidate fixes to the
final merged repair. In total, we recover discussion threads for
approximately \ndisc{} bugs.

\subsubsection{Dataset Representation}

Each dataset instance is centered on a single \syzbot{} bug and contains linked artifacts from three categories:

\begin{itemize}[nosep,leftmargin=*]
    \item \textbf{Bug and crash metadata:} bug ID, title, status, crash timestamps, crash reports, and reproducer artifacts.
    \item \textbf{Repair discussion:} mailing-list messages, thread structure, and patch revision metadata extracted from submission emails.
    \item \textbf{Merged fix:} fixing commit metadata and the final patch diff from the kernel repository.
\end{itemize}

Together, these artifacts provide a bug-centered view of kernel repair that connects the reported failure, the discussion and revision of candidate fixes, and the final accepted patch.

\subsubsection{Analysis Framework}
\label{sec:analysis_framework}

To systematically study \syzbot{}-reported Linux kernel crashes and their corresponding fixes, we built a modular and extensible analysis framework that encodes our measurement pipeline as a set of reusable analyzers. Each analyzer captures one dimension of the repair process, for example, bug type, fix pattern, patch revision behavior, or downstream backporting, and produces structured outputs that can be aggregated across the dataset.
This design allows us to extend the framework with new analyses without changing the underlying data model, and helps ensure that our empirical results are derived through a uniform and reproducible workflow.

At a high level, each analyzer is implemented as a subclass of \texttt{BaseAnalyzer} (Listing~\ref{lst:analyzer}). It consumes a collection of \texttt{BugEntry} objects, where each entry represents a reconstructed bug-fix instance, and returns an \texttt{AnalysisResult} containing aggregate statistics and per-instance records.
This abstraction lets us express different analyses in a consistent form while preserving extensibility
for future studies.

\begin{figure}[H]
\begin{pycode}[title={\small\sffamily\bfseries Python}]
class BaseAnalyzer(ABC):
    @property
    @abstractmethod
    def name(self) -> str: ...

    @abstractmethod
    def analyze(self, bugs: list[BugEntry]) -> AnalysisResult: ...

@dataclass
class AnalysisResult:
    name: str
    summary: dict[str, Any]
    details: list[dict[str, Any]]
    tables: dict[str, list[dict]]
\end{pycode}
\vspace{-1ex}
\captionof{lstlisting}{Common interface for analyzers in our framework.}
\label{lst:analyzer}
\end{figure}

Table~\ref{tab:analyzers} summarizes the analyzers currently implemented in the framework. Although the framework supports a broad range of measurements, this paper focuses on three major themes: bug classification dimensions (\S\ref{sec:dimensions}), patch evolution dynamics (\S\ref{sec:evolution}), and empirical insights and representative case studies drawn from kernel patch evolution (\S\ref{sec:insights}). Together, these analyzers form the methodological foundation for the empirical results presented in the remainder of the paper.

\begin{table}[t]
\centering
\caption{Analyzers implemented in our analysis framework.}
\label{tab:analyzers}
\footnotesize
\resizebox{\columnwidth}{!}{%
\begin{tabular}{@{}lp{6.0cm}@{}}
\toprule
\textbf{Analyzer} & \textbf{Purpose} \\
\midrule
Bug Type & Classifies crashes into bug categories based on crash reports and failure signatures \\
Fix Pattern & Identifies recurring repair patterns from fixing patches \\
Fix Locality & Measures the distance between failure location and fix location at the function, file, and subsystem levels \\
Difficulty & Estimates repair complexity using a composite difficulty score and stratifies bugs into difficulty tiers \\
\textbf{Revision Reasons} & \textbf{Categorizes why an initial patch required revision based on review feedback and patch history} \\
\textbf{Discussion Lessons} & \textbf{Quantifies reviewer participation, discussion depth, and feedback intensity} \\
\textbf{Patch Diff} & \textbf{Characterizes structural differences across patch versions} \\
\textbf{Non-Functional} & \textbf{Identifies revisions driven by non-functional concerns such as style, maintainability, or performance} \\
Info Sufficiency & Evaluates whether the available debugging information is sufficient to localize and repair the bug \\
\textbf{Patch Evolution} & \textbf{Links reviewer feedback on version $v_N$ to structural changes in version $v_{N+1}$, enabling analysis of revision responsiveness} \\
\textbf{Backport Downstream} & \textbf{Tracks whether and how fixes propagate to stable and LTS kernels, including coverage and lag} \\
Case Study Finder & Ranks bug-fix instances by multi-dimensional interestingness for manual inspection \\
\textbf{Insight Clusters} & \textbf{Cross-references multiple analytical dimensions to derive higher-level empirical insight categories} \\
\bottomrule
\end{tabular}%
}
\end{table}

\vspace{1mm}
\textit{Limitations and Scope.}
Our present dataset is intentionally scoped to bugs that can be anchored to fixed \syzbot{} reports. This choice \textit{favors precision over breadth}.
Compared with mining LKML uniformly, \syzbot{}-grounded collection provides clearer bug identities, more reliable failure evidence, and stronger linkage between reports, patch discussions, and merged commits. At the same time, this scope excludes kernel bugs reported through other channels, such as manual testing, code review, or developer-reported regressions outside the \syzbot{} workflow.

The complete dataset occupies 13\,GB on disk: 11\,GB of processed JSON, 1.5\,GB of raw crawled data, and 210\,MB of training data (used for the fine-tuned patch advisor described in \S\ref{sec:advisor}).
\section{Findings from syzbot-Linked Patch Evolution}
\label{sec:findings}

We next present empirical findings from jointly analyzing patch revision histories and review discussions. Rather than treating kernel repair as a one-shot mapping from crash report to final fix, these findings reveal recurring patterns in how fixes are proposed, revised, and accepted in practice.

\subsection{A Multi-dimensional View of Kernel Patch and Patch Evolution}
\subsubsection{Classification Dimensions}
\label{sec:dimensions}

To characterize the structure of kernel bug fixing, we use the analyzers described above to examine five complementary dimensions: bug type, fix pattern, fix locality, repair complexity, and the reasons driving patch revisions.

\emph{Bug Types.}
We classify \nbugs{} bugs into 20 categories by matching crash-report signatures with a set of regular-expression rules.  Table~\ref{tab:bugtypes}
lists the 10 most frequent categories.  Three classes dominate the
dataset: \emph{warning} (21.1\%), \emph{null-pointer dereference}
(14.5\%), and \emph{use-after-free} (14.3\%).  The distribution also
reveals substantial variation in repair characteristics across bug
classes.  For example, \emph{task-hung} bugs exhibit the longest median
fix time (121.4 days), suggesting that liveness and progress failures
are often difficult to diagnose and validate.  In contrast,
\emph{data-race} bugs are resolved most quickly (15.0 days median), but
their fixes are comparatively large, with a median of 17 modified lines.

\begin{table}[t]
\centering
\caption{Top 10 bug types by frequency.}
\label{tab:bugtypes}
\footnotesize
\resizebox{\columnwidth}{!}{%
\begin{tabular}{@{}lrrrr@{}}
\toprule
\textbf{Bug Type} & \textbf{Count} & \textbf{\%} & \textbf{Med.\ Lines} & \textbf{Med.\ Days} \\
\midrule
warning & 1{,}466 & 21.1 & 7 & 28.6 \\
null-ptr-deref & 1{,}007 & 14.5 & 8 & 26.0 \\
use-after-free & 993 & 14.3 & 9 & 48.4 \\
info-leak & 513 & 7.4 & 4 & 47.8 \\
deadlock & 463 & 6.7 & 13 & 34.8 \\
kernel-bug & 373 & 5.4 & 10 & 39.7 \\
out-of-bounds-read & 348 & 5.0 & 9 & 50.7 \\
memory-leak & 257 & 3.7 & 6 & 27.2 \\
task-hung & 244 & 3.5 & 9 & 121.4 \\
data-race & 209 & 3.0 & 17 & 15.0 \\
\bottomrule
\end{tabular}%
}
\end{table}

\emph{Fix Patterns.}
We further identify 10 recurring fix patterns by jointly analyzing
source-level edit operations and commit-message text.  The most common
pattern is \emph{fix-order} (18.9\%), which captures repairs that
reorder operations to eliminate race windows or restore intended control
flow.  This is followed by \emph{add-init} (15.2\%) and
\emph{add-null-check} (12.4\%).  Overall, 58.8\% of bugs match at least
one fix pattern, with an average of 1.07 matched patterns per bug.
These results suggest that while a substantial fraction of fixes exhibit
reusable repair idioms, many kernel fixes still resist reduction to a
single stereotyped edit template.

\emph{Fix Locality.}
To understand how far a fix is from the observed failure site, we
measure locality at the function, file, directory, and subsystem levels.
Among 4{,}986 analyzable bugs, 77.5\% are fixed in the same file as the
crash site, 9.6\% in the same directory, and 6.1\% elsewhere in the
same subsystem.  However, 3.7\% of bugs (183 cases) require edits in a
\emph{different subsystem} entirely.  At a finer granularity,
same-function repairs account for only 3.1\% of cases.  This result
highlights an important mismatch between failure manifestation and fix
location: even when the repair remains file-local, it is rarely confined
to the exact function in which the crash is reported.

\emph{Difficulty Stratification.}
We estimate repair difficulty using a composite score from 0 to 12 based on six observable properties of each fix: patch size, number of modified files, number of patch revisions, locality between the crash site and the patched code, time-to-fix, and whether a C reproducer is available.
Each factor contributes a small ordinal penalty; larger, less localized, longer-running, and more iterative fixes receive higher scores, while the presence of a reproducer lowers the score. Among the \npatch{} scored bugs, 66.4\% fall into the \emph{easy} tier (0--3), 32.6\% into the \emph{medium} tier (4--7), and 1.0\% into the \emph{hard} tier (8--12). Although rare, hard bugs are substantially more demanding: their median fix spans 74 modified lines across 5 files and requires 249.5 days to resolve. This skewed distribution suggests that most syzbot-reported bugs are relatively localized and straightforward to repair, while a small minority impose disproportionately high repair costs.

\emph{Patch Revision Reasons.}
For the 1{,}099 bugs with multiple patch versions, we classify revision reasons into 12 categories based on reviewer comments and inter-version changes.  The most frequent revision drivers are commit-message issues (27.4\%), correctness problems (25.4\%), API design concerns (21.8\%), race conditions (16.7\%), and incomplete fixes (16.5\%).  Notably, 28.7\% of multi-version bugs are revised without intervening external feedback, indicating that self-correction is common, but still subordinate to community-guided refinement.

\subsubsection{Patch Evolution Dynamics}
\label{sec:evolution}

A key feature of our dataset is that it reconstructs the
\emph{evolutionary chain} of a repair: how reviewer feedback on version $v_N$ relates to the structural changes introduced in version $v_{N+1}$.  We use this capability to analyze all 1{,}600 consecutive $v_N \rightarrow v_{N+1}$ transitions drawn from the 1{,}099 bugs with multi-version patch series.

Among the 1,600 version transitions, 1,252 (78.3\%) contain substantive reviewer feedback after excluding bot messages, stable-backport threads, and trivial tag-only replies.  Only 21.7\% of transitions occur without such feedback.  This result shows that patch refinement in the \syzbot ecosystem is primarily a social and collaborative process rather than a sequence of isolated self-revisions.

Table~\ref{tab:evolution} summarizes the 12 feedback categories, ranked by frequency, together with their average structural impact, measured as the absolute number of changed lines between adjacent versions.
Two findings stand out.

\begin{enumerate}[leftmargin=*,nosep]
\item \textbf{The most frequent feedback is not the most disruptive.}
Correctness-related feedback is the most common category (42.1\% of
transitions), followed closely by commit-message feedback (40.2\%).
However, these categories induce relatively small structural changes,
with average impacts of 28.6 and 26.0 lines, respectively.  By
contrast, \emph{performance} feedback appears in only 12.9\% of
transitions but causes the largest average revision footprint (52.8
lines).  Similarly, \emph{style/convention}, \emph{race-condition}, and
\emph{error-handling} feedback all lead to substantially larger
inter-version edits than their frequency alone would suggest.  In other
words, what reviewers mention most often is not necessarily what forces
developers to restructure a patch most extensively.

\item \textbf{Explicit changelogs capture only a small portion of actual
revision behavior.}
Only 36.4\% of transitions include any explicit ``Changes since $v_N$''
notes.  Moreover, changelog acknowledgment is weakly aligned with the
actual categories driving structural revision.  For example,
\emph{commit\_message} issues are relatively likely to be documented
(20.7\%), whereas \emph{performance} feedback is almost never mentioned
in changelogs (1.9\%) despite producing the largest code changes.  The
average cross-category responsiveness score is only 0.072, indicating
that most revision activity remains \emph{implicit}: the patch changes
substantially from one version to the next, but the developer often does
not explicitly record which feedback prompted those changes.
\end{enumerate}

Taken together, these results suggest that patch evolution in the Linux
kernel cannot be understood solely from final merged fixes or even from
explicit changelog annotations.  The underlying process is driven by
review feedback that varies widely in structural consequence, and much
of that consequence is reflected only in the code deltas between patch
versions.

\begin{table}[t]
\centering
\caption{Patch evolution across 1{,}600 version transitions.  ``$\Delta$lines''
denotes the average absolute line-count difference between adjacent patch
versions when the feedback category is present.  ``Changelog'' denotes
the fraction of such transitions in which the category is explicitly
acknowledged in the submission notes of $v_{N+1}$.}
\label{tab:evolution}
\footnotesize
\resizebox{\columnwidth}{!}{%
\begin{tabular}{@{}lrrrr@{}}
\toprule
\textbf{Category} & \textbf{\# Trans.} & \textbf{\%} & \textbf{Avg $\Delta$lines} & \textbf{Changelog} \\
\midrule
correctness      & 673 & 42.1 & 28.6 & 4.5\% \\
commit\_message  & 644 & 40.2 & 26.0 & 20.7\% \\
api\_design      & 587 & 36.7 & 38.2 & 14.8\% \\
race\_condition  & 462 & 28.9 & 48.1 & 8.4\%  \\
incomplete\_fix  & 426 & 26.6 & 46.3 & 4.2\%  \\
documentation    & 408 & 25.5 & 36.2 & 11.0\% \\
config\_build    & 388 & 24.2 & 44.2 & 7.0\%  \\
style\_convention& 376 & 23.5 & 51.3 & 12.8\% \\
error\_handling  & 348 & 21.8 & 46.6 & 9.2\%  \\
scope            & 340 & 21.2 & 32.1 & 4.4\%  \\
memory\_safety   & 305 & 19.1 & 34.5 & 4.9\%  \\
performance      & 206 & 12.9 & \textbf{52.8} & 1.9\%  \\
\bottomrule
\end{tabular}%
}
\end{table}

\subsection{Empirical Insights from Kernel Patch Evolution}
\label{sec:insights}
We next distill our quantitative analysis into a small set of empirical insights that expose recurring properties of Linux kernel repair. Rather than focusing on isolated statistics, these insights highlight the structural traits that make kernel bug fixing distinct from conventional one-shot repair settings.

\subsubsection{Accepted Fixes Are Often Non-Local}
\label{sec:nonlocal}

The crash site is often not the true repair site. Among 4{,}986
analyzable bugs, only 3.1\% are fixed in the same function as the crash.
While 77.5\% are fixed in the same file, a non-trivial fraction require
changes outside the immediate crash context, including 3.7\% whose fixes
reside in a different subsystem entirely. This locality gap suggests
that the stack trace often identifies where a failure manifests, but not
where the accepted repair must be applied.

\noindent\textbf{Example: a kcov warning fixed in the NFC subsystem.}
Bug \texttt{0438378d} triggered a warning in \texttt{kcov\_remote\_start()}, which at first glance appears to implicate the
\texttt{kcov} subsystem~\cite{kcov}. \texttt{kcov} is the kernel's code-coverage facility, primarily used by fuzzing tools to track which kernel paths have been exercised. As a result, a warning at this site can easily be misread as a defect in \texttt{kcov} itself. In reality, however, the accepted fix was not in \texttt{kcov}, but a one-line change in the NFC receive path:

\begin{figure}[H]
\begin{diffcode}[title={\small\sffamily\bfseries net/nfc/nci/core.c}, borderline west={2pt}{0pt}{leftbar}]
diff --git a/net/nfc/nci/core.c b/net/nfc/nci/core.c
--- a/net/nfc/nci/core.c
+++ b/net/nfc/nci/core.c
@@ -1518,6 +1518,7 @@ static void nci_rx_work(...)
        if (!nci_plen(skb->data)) {
            kfree_skb(skb);
+           kcov_remote_stop();
            break;
        }
\end{diffcode}
\vspace{-1ex}
\captionof{lstlisting}{A warning reported in kcov is fixed by adding cleanup in the NFC subsystem.}
\label{lst:nfc-kcov}
\end{figure}

The reported failure occurred in kcov, but the accepted repair restores a missing cleanup step in NFC before control leaves the receive loop (as shown in Listing~\ref{lst:nfc-kcov}).
The bug therefore cannot be understood purely as a local kcov failure.
It is instead an interaction bug in which one subsystem leaves another in an inconsistent state. This kind of symptom--fix mismatch is exactly the kind of signal that is lost when only the final fixing commit is retained.

\subsubsection{Small Final Patches Can Take Long Diagnostic Effort}
\label{sec:smallpatch}

Patch size is a poor proxy for repair difficulty. In our data, many
final fixes are very small, yet took months or even years to converge.
This is especially important for automated repair: generating a short
patch is easy; determining \emph{which} short patch is acceptable is the
hard part.

\noindent\textbf{Example: a two-line HFS+ fix after years of revision.}
Bug \texttt{1c8ff72d}, associated with a warning in
\texttt{hfsplus\_bnode\_create()}, eventually converged to a tiny fix,
but unexpectedly after 5 patch iterations, 32 review messages, and 1{,}166 days
of discussion.
The v1 patch was 76 lines across 3 files; by v2, it
had collapsed to 1 line in 1 file.  The difficulty was not in writing
code but in understanding the complex HFS+ filesystem bnode allocation
logic well enough to identify the correct two-line initialization fix.
The final patch is small, but the repair process reflects
substantial diagnostic difficulty:

\begin{figure}[H]
\begin{diffcode}[title={\small\sffamily\bfseries fs/hfsplus/bnode.c}, borderline west={2pt}{0pt}{leftbar}]
diff --git a/fs/hfsplus/bnode.c b/fs/hfsplus/bnode.c
--- a/fs/hfsplus/bnode.c
+++ b/fs/hfsplus/bnode.c
@@ -... +... @@
-   /* missing initialization */
+   node->page_offset = 0;
+   node->tree = tree;
\end{diffcode}
\vspace{-1ex}
\captionof{lstlisting}{A tiny accepted patch can hide a long and difficult diagnostic process.}
\label{lst:hfs-small-fix}
\end{figure}

What makes this case important is not the number of changed lines, but the contrast between the tiny final repair and the lengthy revision history that preceded it. The final accepted patch does not reveal how many alternate diagnoses were considered and discarded before reviewers and developers converged on the right explanation.

\subsubsection{Reviewer Feedback Encodes Hidden Acceptance Constraints}
\label{sec:constraints}

Review comments frequently raise issues that are not explicit in the
original syzbot report. Across revision transitions, common feedback
categories include correctness, API design, race conditions,
incomplete-fix concerns, error handling, build compatibility, and
performance. Moreover, high-impact feedback is not always the most
frequent: \textit{performance concerns appear less often than correctness
comments}, yet they trigger \textit{some of the largest structural changes across
versions}.

This shows that patch acceptance depends on more than suppressing the
reported crash. Maintainers evaluate whether a proposed change respects
subsystem invariants, locking rules, publication order, and interaction
protocols with neighboring components.

\noindent\textbf{Example: locking and publication order in L2TP.}
Bug \texttt{94cc2a66} was reported as a locking issue in
\texttt{inet\_autobind()}, but the accepted fix was in
\texttt{net/l2tp/l2tp\_core.c}, where the tunnel was being published
before all socket state had been safely initialized:

\begin{figure}[H]
\begin{diffcode}[title={\small\sffamily\bfseries net/l2tp/l2tp\_core.c}, borderline west={2pt}{0pt}{leftbar}]
diff --git a/net/l2tp/l2tp_core.c b/net/l2tp/l2tp_core.c
--- a/net/l2tp/l2tp_core.c
+++ b/net/l2tp/l2tp_core.c
@@ -1502,6 +1502,10 @@ int l2tp_tunnel_register(...)
+    lock_sock(sk);
+    write_lock_bh(&sk->sk_callback_lock);
     /* ... initialize socket fields ... */
+    write_unlock_bh(&sk->sk_callback_lock);
+    release_sock(sk);
+
     /* Publish the tunnel -- AFTER initialization */
     ...
\end{diffcode}
\vspace{-1ex}
\captionof{lstlisting}{The accepted fix enforces locking and initialization order before publication.}
\label{lst:l2tp-locking}
\end{figure}

This example illustrates two important points. First, the accepted fix
is non-local: the warning surfaced in one place, but the repair changed
ordering and synchronization elsewhere. Second, the final patch is
shaped by constraints that are only visible through review discussion:
the code must not merely avoid the observed warning, but must also
satisfy socket-locking discipline and publication-order invariants.

Taken together, these findings suggest that kernel repair should not be
modeled as a direct mapping from crash report to final patch. In many
syzbot-linked LKML threads, the initial candidate patch is revised
multiple times, reviewer feedback introduces constraints not stated in
the bug report, and the accepted fix may differ substantially from the
apparent crash location. Patch evolution history therefore provides a
richer supervision signal than final fixes alone: it records not only
what was eventually merged, but also what earlier versions got wrong and
how maintainers guided the repair toward an acceptable solution.
\section{From Patch-Evolution Insights to Actionable Repair Methods}
\label{sec:methods}

The analysis in \S\ref{sec:findings} establishes that syzbot crash
repair cannot be adequately captured as a direct crash-to-patch mapping.
Accepted fixes are often non-local, even small final patches may reflect
substantial diagnostic effort, and reviewer feedback frequently imposes
acceptance constraints absent from the original crash report. These
findings motivate a repair system that goes beyond code generation:
it should identify likely root-cause directions, expose subsystem-specific
constraints, and steer patch generation toward solutions consistent with
historically accepted repairs.

Guided by these observations, we design a two-part repair method.
First, we build a \emph{memory layer} that distills our syzbot-linked
patch-evolution dataset into retrievable bug instances, fix patterns,
and review lessons. Second, we fine-tune a lightweight
\emph{diagnostic model} to generate a repair-oriented summary from
a new crash report and an initial candidate patch. At inference
time, the memory layer provides historical context, while
the fine-tuned model synthesizes a compact diagnosis that guides a
general-purpose coding agent toward a reviewer-aligned patch.

\subsection{Patch Evolution Memory}
\label{sec:evolution_memory}

The memory component is built directly from the reconstructed syzbot-linked dataset and its analyzer outputs. Its goal is not to memorize coding details, but to expose historically useful repair context at inference time.

The memory has three layers.
The first layer is \textit{instance memory}, where each bug is represented as a structured \texttt{BugMemoryEntry} containing: (1) crash evidence such as bug type, stack frames, and implicated files; (2) fix characterization, including repair pattern, locality, and difficulty tier; (3) the final merged diff and commit message; and (4) a compact patch-evolution summary, including revision count, major revision reasons, and key reviewer feedback.

The second layer is \textit{pattern memory}, which aggregates individual bug records into reusable repair knowledge.
For each observed \texttt{(bug\_type, fix\_pattern)} pair, we store a \texttt{FixStrategy} record with frequency, median patch size, representative examples, and recurring pitfalls. We also build \texttt{ReviewLesson} records from revision-reason categories to capture common pre-merge concerns, such as incomplete fix, race-condition handling, API misuse, and error-path cleanup.
The third layer is \textit{embedding memory}. To support semantic retrieval, we map crash reports and patch artifacts into a dense vector space, indexing them for nearest-neighbor search. This layer enables the system to retrieve historically similar bugs even when their textual surface forms, such as variable names or specific error messages, differ significantly.

Given a new crash report, the retriever performs four steps:

\begin{enumerate}[leftmargin=*,nosep,label=\arabic*.]
  \item classify the likely bug type from the report text;
  \item retrieve semantically similar historical bugs;
  \item collect high-frequency fix strategies associated with the
  predicted bug type and retrieved neighbors; and
  \item surface review lessons associated with likely fix patterns and
  known failure modes.
\end{enumerate}

The retrieved result is formatted into a bounded context block for
the downstream advisor. In practice, this context includes: similar
bugs, likely repair directions, warnings about common traps, and
subsystem-specific review concerns.

The memory layer directly addresses the non-locality and acceptance constraint problems identified in \S\ref{sec:findings}.
For example, when a crash is likely to be a misleading symptom of
state corruption elsewhere, retrieval can surface historical examples
whose accepted fixes occur away from the apparent crash site.
Likewise, when a bug pattern frequently triggers review concerns
about ordering, cleanup, or synchronization, these lessons can be
made explicit before code generation begins.

\subsection{Fine-Tuned Diagnostic Patch Advisor}
\label{sec:advisor}

While memory retrieves relevant precedent, it does not by itself produce a coherent diagnosis for a new bug. We therefore fine-tuned a small advisor model to synthesize a repair-oriented summary that bridges crash evidence, candidate patch content, and retrieved historical context.

Raw discussions are valuable but noisy. Instead of training directly on thread text, we derive a cleaner supervision target from patch evolution. For each multi-version patch series, we construct a diagnostic target that summarizes three aspects of the repair process: the logical gap between the initial candidate patch and the final accepted repair, the maintainer concerns expressed during revision, and the high-level repair strategy that resolves the bug without violating subsystem constraints.

This target is intentionally \emph{code-free}: the advisor is trained
to explain what kind of fix is needed, not to emit the final patch.
This separation keeps diagnosis focused on architectural and review
constraints rather than token-level code imitation.

\subsubsection{Model Objective}

Let $X$ denote the input context for a training instance, consisting
of the crash report, the initial candidate patch, and the retrieved
memory context. Let $Z$ denote the target diagnostic summary. We
fine-tune the advisor with a standard conditional language-model
objective:
\[
\mathcal{L}_{\text{advisor}}
=
-\sum_{k=1}^{|Z|}
\log P(z_k \mid X, z_{<k}).
\]
Unlike direct patch-generation training, this objective teaches the
model to predict \emph{repair guidance} rather than final code. In
effect, the advisor learns to approximate the kinds of critiques and
constraints that maintainers apply during patch refinement.

\subsubsection{Training Setup}

We construct the fine-tuning corpus from \nbugs{} syzbot-fixed bugs, filtering out instances that have only a single patch version or whose fixing commit is missing from the kernel tree. Each remaining multi-version patch series yields one or more input--output training pairs, combining crash reports, initial patches, and memory context as input with diagnostic summaries as targets. We fine-tune two base models---Gemma-3-12B and LLaMA-3.2-11B---using LoRA on a single NVIDIA A100 80\,GB GPU.

\subsubsection{Advisor Output Format}

For a new crash, the advisor emits a concise diagnostic summary containing four fields: \textit{likely root-cause direction}, which indicates whether the apparent crash site is likely to be the true repair site; \textit{repair constraints}, which capture requirements such as locking, lifetime, ordering, cleanup, build compatibility, or API usage; \textit{patch strategy}, which describes the expected kind of fix, such as reordering operations, restoring missing cleanup, adding initialization, or narrowing the publication window; and \textit{pitfalls to avoid}, which are distilled from historical review feedback.
This representation is compact enough for prompting, but specific enough to constrain the downstream coder.

\subsubsection{Patch Generation with Advisor Guidance}

The final patch is generated by a general-purpose coding model, which we treat as a \emph{coder}. The coder receives a composite context consisting of the crash report, the candidate patch (when available), the retrieved memory context, and the advisor summary.
We instruct the coder to treat the advisor output as a repair specification rather than optional prose.
To improve syntactic robustness, we wrap generation in a compilation-feedback loop. After a patch is produced, we attempt to apply it and compile the affected subsystem. If compilation fails, the compiler output is appended to the prompt and the coder is asked to revise the patch. This process filters out syntactically invalid candidates before semantic evaluation.

\subsection{Evaluation}
\label{sec:eval}

We reserve a temporally held-out evaluation set of 100 previously unseen crash cases, excluded from both memory construction and model training.
To evaluate memory utility, we provide each crash report and its initial patch (v1) as input and ask the model to suggest improvements likely to be raised during review.
Ground-truth labels are derived by mapping human reviewer comments from LKML to the 12 revision categories (Table~\ref{tab:evolution}) using the same keyword-matching procedure as dataset construction; the complete keyword lists are included in the accompanying artifact.
Enabling memory consistently improves alignment with actual reviewer feedback: F1 increases by \textbf{21.8\%} (0.330 vs.\ 0.271), with precision up \textbf{20.7\%} and recall up \textbf{16.7\%}, suggesting that prior repair history helps the model anticipate the concerns reviewers are likely to raise.

We next evaluate \emph{end-to-end patch generation quality} on recent \syzbot crash cases under three configurations:
(1) \textit{Baseline} (coding agent receives only the crash report and candidate patch),
(2) \textit{Memory-guided} (agent additionally receives retrieved memory context), and
(3) \textit{Advisor-guided} (agent receives both memory context and the fine-tuned advisor's diagnostic summary).

\paragraph{Bug selection criteria.}
To ensure temporal separation from the fine-tuning data, we restrict the evaluation set to recently fixed bugs that were \emph{not} included in the advisor training corpus.
From the 30 most recently fixed syzbot bugs at the time of evaluation, we apply three selection criteria: (1)~a working C reproducer must be available, (2)~the patch must have undergone at least two revision iterations on LKML (indicating non-trivial reviewer interaction), and (3)~the bug must not appear in the fine-tuning dataset.
Only 6 of the 30 candidates satisfy all three conditions.

\paragraph{Preventing evaluation leakage.}
Because all six bugs were already fixed upstream, we explicitly \emph{prohibit} the coding agent from querying Linux kernel commits, LKML archives, or the syzbot dashboard via prompt-level instructions, applied uniformly across all configurations.

We report both diagnostic and end-to-end repair quality in Tables~\ref{tab:fix_outcomes} and~\ref{tab:aggregate_metrics}.
For diagnosis, \emph{Diagnostic Coverage Rate (DCR)} measures the proportion of ground-truth revision categories (from subsequent human review) covered by the generated summary.
For end-to-end repair, we report CodeBERTScore~\cite{codebertscore:github,codebertscore} (semantic similarity to the reference patch via CodeBERT embeddings), compilation success rate, and Avg.\ Rank (lower is better), computed by presenting all anonymized patches for the same bug to Claude Sonnet 4.6 in random order and averaging the resulting quality ranks across bugs.
Each patch is applied at the exact bug-introducing commit and verified via kernel compilation and 600-second QEMU/KVM reproducer execution on a dual-socket Intel Xeon Platinum 8380 server (80 cores, 125\,GB RAM) running Ubuntu 22.04 with QEMU 6.2.0.

\begin{table}[t]
    \centering
    \caption{Per-bug fix outcomes on 6 \syzbot bugs (February 2026). \ding{51} = compiled \& runtime-verified fix; \ding{51}$^*$ = patch equivalent to ground truth under GCC toolchain limitation; \ding{55} = incorrect or non-equivalent fix. All configurations use the same coder (Claude). ``Base model'' denotes the unfine-tuned advisor.}
    \label{tab:fix_outcomes}
    \small
    \setlength{\tabcolsep}{2.5pt}
    \begin{tabular}{l cccccc c}
        \toprule
        \textit{Configuration} & \textit{48dc} & \textit{4e2e} & \textit{56b6} & \textit{827a} & \textit{d00f} & \textit{f500} & \textit{Fix Rate} \\
        \midrule
        Baseline & \ding{55} & \ding{55} & \ding{51} & \ding{55} & \ding{55} & \ding{55} & 1/6 \\
        Memory-guided & \ding{51} & \ding{51}$^*$ & \ding{51} & \ding{51} & \ding{55} & \ding{55} & 4/6 \\
        \midrule
        \multicolumn{8}{l}{\textit{Base model Advisor (No Fine-tuning)}} \\
        LLaMA-3.2-11B (Base) & \ding{51} & \ding{51}$^*$ & \ding{51} & \ding{51} & \ding{55} & \ding{55} & 4/6 \\
        Gemma-3-12B (Base) & \ding{51} & \ding{51}$^*$ & \ding{51} & \ding{51} & \ding{55} & \ding{55} & 4/6 \\
        \midrule
        \multicolumn{8}{l}{\textit{Advisor-guided (\sys, Fine-tuned)}} \\
        \textbf{Gemma-3-12B} & \ding{51} & \ding{51}$^*$ & \ding{51} & \ding{51} & \ding{51} & \ding{55} & \textbf{5/6} \\
        LLaMA-3.2-11B & \ding{51} & \ding{55} & \ding{51} & \ding{51} & \ding{55} & \ding{55} & 3/6 \\
        \bottomrule
    \end{tabular}
    \vspace{1ex}
    \\ \small \textit{Note: $^*$Bugs 4e2e and f500 ground truth patches are also affected by GCC vs.\ Clang compilation differences.}
\end{table}

\begin{table}[t]
    \centering
    \caption{Aggregate quality metrics across all configurations. DCR = Diagnostic Coverage Rate; CodeBERT = CodeBERTScore; Avg.\ Rank = mean AI-assessed quality ranking ($\downarrow$ lower is better). Baseline and Memory-guided lack a diagnostic step; DCR is not applicable (--).}
    \label{tab:aggregate_metrics}
    \small
    \begin{tabular}{l cccc}
        \toprule
        \textit{Configuration} & \textit{DCR} & \textit{CodeBERT} & \textit{Compile} & \textit{Avg.\ Rank$\downarrow$} \\
        \midrule
        Baseline & -- & 0.42 & 83.3\% & 5.2 \\
        Memory-guided & -- & 0.74 & 100.0\% & 3.2 \\
        \midrule
        LLaMA-3.2-11B (Base) & 0.62 & 0.76 & 100.0\% & 3.2 \\
        Gemma-3-12B (Base) & 0.65 & 0.78 & 100.0\% & 3.2 \\
        \midrule
        \textbf{Gemma-3-12B (FT)} & \textbf{0.88} & \textbf{0.91} & \textbf{100.0\%} & \textbf{2.7} \\
        LLaMA-3.2-11B (FT) & 0.71 & 0.68 & 100.0\% & 3.7 \\
        \bottomrule
    \end{tabular}
\end{table}

Tables~\ref{tab:fix_outcomes} and~\ref{tab:aggregate_metrics} show a clear monotonic improvement across configurations. The baseline agent succeeds on only 1 of 6 bugs; adding memory retrieval raises this to 4 of 6, indicating that historical repair precedents already provide substantial value. The fine-tuned Gemma-3-12B advisor-guided configuration achieves the best overall results: 5 of 6 bugs fixed, the highest CodeBERTScore (0.91), 100\% compilation rate, and the best average quality rank (2.7). Notably, it is the only configuration to correctly resolve the challenging \texttt{d00f} case, whose fix requires cross-function causal reasoning---all other configurations target the wrong code location. In contrast, the fine-tuned LLaMA-3.2-11B advisor achieves only 3/6, below its own base model (4/6), suggesting that fine-tuning on evolutionary traces can overfit to surface patterns when model capacity is limited. This underscores that fine-tuning benefits are model-dependent. Bug f500 (ATM subsystem) remains unsolved by all configurations, as it requires multi-file RCU refactoring beyond current model capabilities. Overall, the key benefit comes from making latent repair constraints explicit before generation: memory retrieves precedents, and the advisor converts them into bug-specific guidance.

\begin{figure}[t]
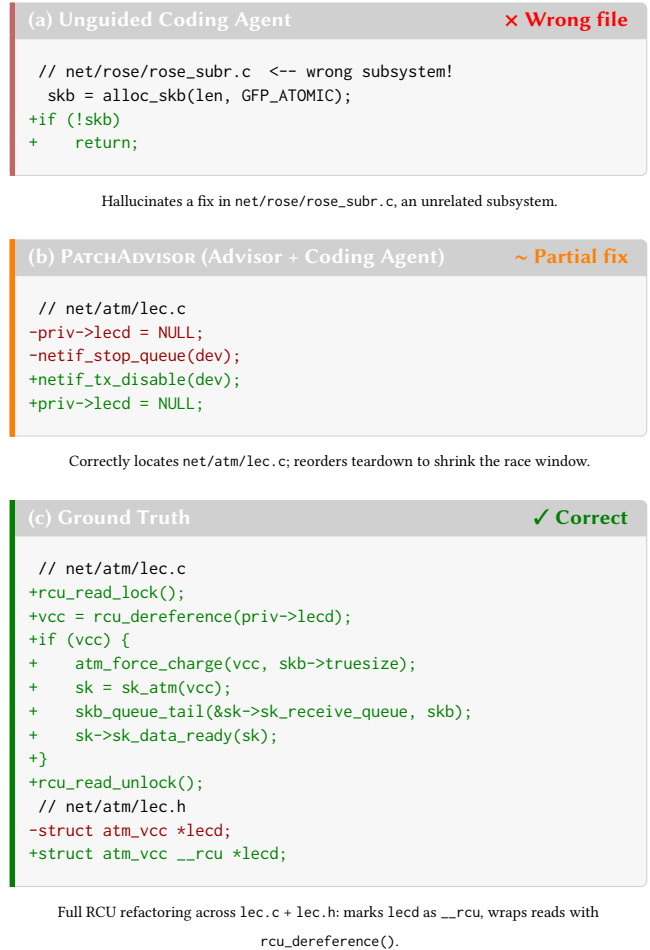

    \centering

    \begin{diffcode}[title={\small\sffamily\bfseries (a) Unguided Coding Agent \hfill \textcolor{red}{$\boldsymbol\times$ Wrong file}}]
 // net/rose/rose_subr.c  <-- wrong subsystem!
  skb = alloc_skb(len, GFP_ATOMIC);
+if (!skb)
+    return;
    \end{diffcode}
    \vspace{-2pt}
    {\scriptsize Hallucinates a fix in \texttt{net/rose/rose\_subr.c}, an unrelated subsystem.}

    \vspace{4pt}

    \begin{diffcode}[title={\small\sffamily\bfseries (b) \sys (Advisor + Coding Agent) \hfill \textcolor{orange}{$\boldsymbol\sim$ Partial fix}}, borderline west={2pt}{0pt}{orange}]
 // net/atm/lec.c
-priv->lecd = NULL;
-netif_stop_queue(dev);
+netif_tx_disable(dev);
+priv->lecd = NULL;
    \end{diffcode}
    \vspace{-2pt}
    {\scriptsize Correctly locates \texttt{net/atm/lec.c}; reorders teardown to shrink the race window.}

    \vspace{4pt}

    \begin{diffcode}[title={\small\sffamily\bfseries (c) Ground Truth \hfill \textcolor{green!50!black}{\ding{51} Correct}}, borderline west={2pt}{0pt}{green!50!black}]
 // net/atm/lec.c
+rcu_read_lock();
+vcc = rcu_dereference(priv->lecd);
+if (vcc) {
+    atm_force_charge(vcc, skb->truesize);
+    sk = sk_atm(vcc);
+    skb_queue_tail(&sk->sk_receive_queue, skb);
+    sk->sk_data_ready(sk);
+}
+rcu_read_unlock();
 // net/atm/lec.h
-struct atm_vcc *lecd;
+struct atm_vcc __rcu *lecd;
    \end{diffcode}
    \vspace{-2pt}
    {\scriptsize Full RCU refactoring across \texttt{lec.c} + \texttt{lec.h}: marks \texttt{lecd} as \texttt{\_\_rcu}, wraps reads with \texttt{rcu\_dereference()}.}

    \caption{Patch comparison for \textbf{KASAN slab-use-after-free in \texttt{sock\_def\_readable}} (bug f5007221, ATM LEC subsystem). (a)~Without guidance, the coding agent targets the wrong subsystem. (b)~With advisor context, it locates the correct file. (c)~The official fix requires a full RCU refactoring beyond current model capabilities.}
    \label{fig:code_compare}
\end{figure}

\subsection{Threats to Validity}
\label{sec:threats}

\emph{External validity.}
The end-to-end evaluation covers only 6 bugs due to strict selection criteria (working reproducer, multi-version history, temporal separation from training data). While we report per-bug outcomes transparently and use hedged language throughout, this sample size precludes statistical significance testing. We view the end-to-end results as a proof-of-concept demonstration rather than a definitive benchmark; scaling to a larger evaluation set is a priority for future work. The memory evaluation on 100 held-out cases provides broader, though less end-to-end, evidence.

\emph{Internal validity.}
Evaluation leakage is mitigated via prompt-level instructions that prohibit the agent from querying upstream commits, LKML archives, or the syzbot dashboard. Beyond log inspection, we manually reviewed the agent's chain-of-thought reasoning for all runs to confirm that no ground-truth patch content was retrieved or reproduced.
All six evaluation bugs were fixed after the knowledge cutoff dates of both the coding agent and the advisor base models, so their fixes do not appear in any pre-training corpus. The chain-of-thought analysis further confirmed that the agent's reasoning derived from the provided crash reports and advisor summaries rather than recalling known patches.

\emph{Construct validity.}
Avg.\ Rank relies on an LLM judge (Claude Sonnet 4.6), which may not perfectly align with human expert assessment. We mitigate this by presenting patches anonymously in randomized order and prompting for structured ranking criteria. CodeBERTScore may undervalue functionally correct patches that differ structurally from the reference; we therefore report it alongside fix rate and compilation rate rather than as a standalone metric. Our evaluation uses a single coding agent (Claude); results may vary with other code generation models, though our framework is agent-agnostic by design.

\section{Related Work}
\label{sec:related}

\emph{Bug-fix and vulnerability datasets.}
A large body of prior work has built datasets for software testing, debugging, and repair.
Defects4J established a widely used benchmark of real Java bugs with reproducible faulty and fixed versions plus triggering tests, making controlled APR evaluation practical~\cite{just2014defects4j}.
BugsInPy extended this idea to Python, providing real bugs from open-source projects for testing and debugging research~\cite{widyasari2020bugsinpy}.
In the security domain, CVEfixes links public vulnerability records to source-level fixes~\cite{bhandari2021cvefixes}, while MegaVul scales vulnerability mining for C/C++ code and provides richer code representations for learning-based analysis~\cite{ni2024megavul}. Also, the ARVO dataset~\cite{Arvo-2024}, which consists of thousands of Docker containerized images, is used to reproduce OSS-Fuzz~\cite{OSS-Fuzz} vulnerabilities in C/C++, along with their corresponding fixes.
Compared with these datasets, our focus is on \emph{Linux kernel bugs reported by syzbot} and, more importantly, on capturing the \emph{full repair lifecycle}---including crash reports, reproducers, patch revisions, and review discussions---rather than only a buggy/fixed snapshot.

\emph{LLM-based automated program repair.}
Recent work on APR increasingly uses large language models as repair engines.
ChatRepair showed that conversational repair, which iteratively incorporates validation feedback, can substantially improve over one-shot prompting~\cite{xia2023chatrepair}.
RepairAgent further pushed this direction by framing APR as an autonomous agent workflow that gathers information, invokes tools, and iteratively refines candidate fixes~\cite{Bouzenia-APR-LLM-2025}.
SWE-bench broadened evaluation from synthetic or narrowly scoped bugs to real GitHub issues, showing that realistic software engineering repair remains difficult for contemporary models~\cite{jimenez2024swebench}. Furthermore, APPATCH uses union slicing, assisted by semantic matching, to narrow the scope of vulnerabilities' code-related context to support the model's reasoning~\cite{Nong-APR-LLM-2025}. On the other hand, PatchAgent~\cite{Yu-APR-LLM-2025} uses the Language Server Protocol (LSP)~\cite{LSP} to allow structured code retrieval and analysis within a repository.
However, these lines of work mainly target user-space repositories and rely on issue trackers, tests, or validation loops that do not reflect the Linux kernel's email-driven review process, subsystem structure, concurrency hazards, and crash-centric debugging workflow.
Our work complements them by modeling \emph{kernel-native repair artifacts} and by exposing supervision for tasks beyond patch generation, including review, revision, and preference learning.

\emph{Kernel-specific repair benchmarks and environments.}
The closest kernel-specific benchmark is KGym, which introduces an execution platform and benchmark for Linux kernel crash resolution using real bugs, crash traces, reproducers, and developer fixes~\cite{mathai2024kgym}.
Its baseline results also highlight how poorly current LLMs perform on kernel repair tasks.
Our work is complementary to KGym: KGym emphasizes \emph{end-to-end evaluation in an executable environment}, whereas our dataset emphasizes \emph{large-scale lifecycle mining} and \emph{supervision from patch review and revision histories}.
In this sense, our dataset can serve as a training and analysis resource, while KGym provides a complementary execution-based evaluation perspective.

\emph{Kernel patch correctness, evolution, and patch ecosystems.}
Beyond patch generation itself, prior work has shown that kernel patches are often subtle, evolving, and error-prone artifacts.
KLAUS studies incorrectly developed Linux kernel patches and proposes a method to evaluate patch correctness, showing that even accepted patches can introduce serious security risks~\cite{wu2023klaus}.
At a broader ecosystem level, studies of the Android kernel patch ecosystem show that propagation of upstream kernel patches can be delayed by months or years, underscoring the complexity of patch management and the difficulty of identifying security-critical upstream commits~\cite{zhang2021androidkernelpatch}.
Recent work further highlights that security patches do not remain static after their initial landing.
Xie et al.~\cite{xie2024unveiling} present a systematic study of security patch evolution over 1,046 CVEs and 2,633 patches from open-source projects, including Linux, and show that post-fix evolution is both common and consequential for security analysis.
These observations align with our motivation, but the focus is different.
While prior work studies patch correctness or security patch evolution at a broad ecosystem level, our work centers on the repair lifecycle of syzbot-reported Linux kernel bugs and captures finer-grained supervision signals, including crash reports, reproducers, revision chains, and reviewer feedback.
Together, these studies reinforce that, in kernel development, producing a patch is only one step; correctness, revision, review, and downstream impact all matter.
Our work contributes a dataset that makes these process signals explicit and learnable.

\emph{Empirical studies of syzbot and continuous kernel fuzzing.}
Continuous fuzzing infrastructure such as syzkaller/syzbot has become a major source of Linux kernel bug discovery.
Empirical work on continuous kernel fuzzing observed large numbers of unresolved crashes, characterized common bug types, and found that only a subset of fixes went through substantial review or additional testing~\cite{ruohonen2019continuouskernelfuzzing}.
Our work builds on this line of research by moving from aggregate measurement to \emph{instance-level repair lifecycle reconstruction}, connecting each bug report to reproducers, final fixes, patch iterations, and reviewer feedback.
This enables not only empirical analysis of what makes bugs difficult, but also the construction of training tasks tailored to post-training LLMs for kernel repair.

\section{Conclusion}
\label{sec:conclusion}

We investigated Linux kernel patch evolution through a large \syzbot{}-grounded dataset, confirming that accepted fixes are frequently iterative, non-local, and shaped by reviewer-enforced constraints. Motivated by these findings, we developed \sys{}, which leverages patch-evolution history to guide AI-assisted repair. Our results indicate that revision history constitutes a valuable supervision signal for kernel repair and that effective automated fixing should model the refinement process rather than treat patch generation as a one-shot task.

Future directions include scaling evaluation to a larger bug set, incorporating multi-file architectural reasoning for complex repairs such as RCU refactoring, and integrating stronger verification for concurrency-sensitive patches.

\bibliographystyle{ACM-Reference-Format}
\bibliography{refs}

\end{document}